\documentclass[10pt,twocolumn,oneside,final]{IEEEtran}

\IEEEoverridecommandlockouts
\usepackage{cite}
\usepackage{graphicx}
\usepackage{amsmath}
\usepackage{times}
\usepackage{latexsym}
\usepackage{bm}
\usepackage{amssymb}
\usepackage[center]{caption2}
\usepackage{array}
\usepackage{fancyhdr}
\usepackage{cite,graphicx,amsmath,amssymb}
\usepackage{citesort}
\usepackage{psfrag}
\usepackage{multirow}
\usepackage{array}
\usepackage{subfigure}
\usepackage{stfloats}
\usepackage{url}
\usepackage{nicefrac}
\usepackage{amssymb}
\usepackage{amsmath}
\usepackage{amsfonts}
\usepackage{amssymb}
\usepackage{float}
\usepackage{kbordermatrix}

\ifCLASSINFOpdf

\else

\fi

\usepackage[dvipsnames,usenames]{color}
\usepackage[usenames,dvipsnames]{color}

    \def\Complex{{\rm\rule[.23ex]{.03em}{1.1ex}\kern-.3em{C}}}

    \newcommand{\be}{\begin{equation}} \newcommand{\ee}{\end{equation}}
    \newcommand{\bea}{\begin{eqnarray}} \newcommand{\eea}{\end{eqnarray}}
    \newcommand{\benum}{\begin{enumerate}} \newcommand{\eenum}{\end{enumerate}}

    \newcommand{\qG}{{\bf G}}

    \newcommand{\tr}{{\sf tr}}

\newtheorem{theorem}{Theorem}

\newtheorem{alg}{Algorithm}
\newtheorem{definition}{Definition}

\begin{document}
\title{Large-Scale MIMO Secure Transmission with Finite Alphabet Inputs}

\author{\IEEEauthorblockN{Yongpeng Wu, Jun-Bo Wang, Jue Wang, Robert Schober, and Chengshan Xiao}


\thanks{Y. Wu and R. Schober are with the Institute for Digital Communications, University Erlangen-N$\ddot{u}$rnberg,
Cauerstrasse 7, D-91058 Erlangen, Germany (Email: yongpeng.wu@fau.de; robert.schober@fau.de).}

\thanks{J.-B. Wang is with the National Mobile Communications Research Laboratory,
Southeast University, Nanjing, 210096, P. R. China (Email: jbwang@seu.edu.cn).}

\thanks{J. Wang is with School of Electronics and Information, Nantong University, Nantong, 226000, P. R. China (Email: wangjue@ntu.edu).}

\thanks{C. Xiao is with the Department of Electrical and Computer Engineering,
Missouri University of Science and Technology, Rolla, MO 65409, USA (Email: xiaoc@mst.edu). }

}
\maketitle

\begin{abstract}
In this paper, we investigate  secure transmission over the large-scale
multiple-antenna wiretap channel with finite alphabet inputs.
First, we show analytically that a generalized singular value decomposition (GSVD)
based design, which is optimal for Gaussian inputs,
may exhibit a severe performance loss for finite alphabet inputs in the high signal-to-noise ratio (SNR) regime.
In light of this, we propose a novel Per-Group-GSVD (PG-GSVD) design
which can effectively compensate the performance loss caused by the GSVD
design. More importantly,  the computational complexity of the PG-GSVD design  is
 by orders of magnitude lower than that of the existing
design for finite alphabet inputs in \cite{Wu2012TVT} while the resulting performance loss is minimal.
Numerical results  indicate that the proposed PG-GSVD design can be efficiently implemented
in large-scale multiple-antenna systems and achieves significant performance gains compared to the GSVD design.
\end{abstract}


\section{Introduction}
Security is a critical issue for future 5G
wireless networks. In today's systems, the security provisioning relies on bit-level cryptographic techniques
and associated processing techniques at various stages of the data protocol stack. However, these solutions have
severe drawbacks and many weaknesses of
standardized protection mechanisms for public wireless networks  are well known;
although enhanced ciphering and authentication protocols
exist, they impose  severe constraints and high additional costs for the users of public wireless networks.
Therefore, new security approaches based on information theoretical considerations
have been proposed and are collectively referred to as physical layer security  \cite{Wyner1975BSTJ,Khisti2010TIT_2,Gursoy2012TCom,Zeng2016TWC,Wu2016TIT}.

Most existing work on physical layer security assumes that the input signals are Gaussian distributed.
Although the Gaussian codebook has been proved to achieve the secrecy capacity of the Gaussian wiretap channel \cite{Khisti2010TIT_2},
the signals employed in practical communication systems are non-Gaussian and are often drawn from
 discrete constellations \cite{Lozano2006TIT,Xiao2011TSP,Wu2012TWC,Wu2013TCOM}. For the multiple-input, multiple-output, multiple antenna eavesdropper (MIMOME) wiretap
channel with perfect channel state information (CSI) of both the desired user and the eavesdropper
at the transmitter, a generalized singular value decomposition (GSVD) based
precoding design was proposed to decouple the corresponding wiretap channel into  independent parallel subchannels \cite{Bashar2012TCom}.
Then, the optimal power allocation policy across these subchannels was obtained by an iterative algorithm.
However, the simulation results in \cite{Wu2012TVT} revealed that for finite alphabet inputs, the GSVD design is suboptimal.
In fact, the iterative algorithm in \cite{Wu2012TVT} can significant improve the
secrecy rate by directly optimizing the precoder matrix.
Very recently, for the case when imperfect CSI of the eavesdropper is available at the transmitter,
a secure transmission  scheme  was proposed in \cite{Aghdam2016}
based on the joint design of the transmit precoder matrix to improve the achievable rate
of the desired user and the AN generation scheme to
degrade the achievable rate of the eavesdropper.
However, the computational complexities  of the algorithms in
 \cite{Wu2012TVT} and \cite{Aghdam2016} scale
exponentially with the number of transmit antennas.
Therefore, the algorithms in \cite{Wu2012TVT,Aghdam2016} become intractable
even for a moderate number of transmit antennas (e.g., eight).

In this paper, we investigate the secure transmission design
for the large-scale MIMOME wiretap channel with finite alphabet inputs and perfect CSI of the desired user and the eavesdropper at the transmitter.
The contributions of our paper are summarized as follows:

\begin{enumerate}

\item We derive an upper bound on the secrecy rate for finite alphabet inputs in the high SNR regime
when the GSVD design is employed.
The derived expression shows that,  when $N_t > N_1$, in the high SNR regime, the GSVD design will result
in at least $(N_t - N_1) \log M$ b/s/Hz rate loss compared to the maximal rate
for the MIMOME wiretap channel, where $N_t $, $N_1$, and $M$ denote
the number of transmit antennas, the rank of the intended receiver's channel, and the size of the input signal constellation set,
respectively.

\item To tackle this issue, we propose a novel Per-Group-GSVD (PG-GSVD)
design, which pairs different subchannels into different groups based on the GSVD structure.
We prove that the proposed PG-GSVD design can eliminate the performance loss of the GSVD design
with an order  of magnitude lower computational complexity than the design in \cite{Wu2012TVT}. Accordingly, we propose an iterative algorithm based on
the gradient descent method to optimize the secrecy rate.

\item Simulation results illustrate that the proposed designs are well suited for large-scale MIMO wiretap channels and
achieve substantially higher secrecy rates than the GSVD design while requiring a much
lower computational complexity than the precoder design in \cite{Wu2012TVT}.

\end{enumerate}

\emph{Notation:}  Vectors are denoted by lower-case bold-face letters;
matrices are denoted by upper-case bold-face letters. Superscripts $(\cdot)^{T}$, $(\cdot)^{*}$, and $(\cdot)^{H}$
stand for the matrix transpose, conjugate, and conjugate-transpose operations, respectively. We use  ${\tr}({\bf{A}})$ and ${\bf{A}}^{-1}$
to denote the trace and the
inverse of matrix $\bf{A}$, respectively.
$^{\bot}$ denotes the orthogonal complement of a subspace.
 ${\rm{diag}}\left\{\bf{b}\right\}$ denotes a diagonal matrix
with the elements of vector $\bf{b}$ on its main diagonal.
${\rm{Diag}}\left\{\bf{B}\right\}$  denotes a diagonal matrix containing in the main diagonal
the diagonal elements of matrix $\mathbf{B}$.
The $M \times M$ identity matrix is denoted
by ${\bf{I}}_M$, and the all-zero $M \times N$ matrix and the all-zero $N \times 1$ vector are denoted by $\bf{0}$.
The fields of complex numbers and real numbers are denoted
by $\mathbb{C}$ and $\mathbb{R}$, respectively. $E\left[\cdot\right]$ denotes statistical
expectation. $[\mathbf{A}]_{mn}$ denotes the element in the
$m$th row and $n$th column of matrix $\mathbf{A}$. $[\mathbf{a}]_{m}$ denotes the $m$th entry
of vector $\mathbf{a}$. We use  $\mathbf{x} \sim \mathcal{CN} \left( {\mathbf{0},{{\bf{R}}_N}} \right)$
to denote a circularly symmetric complex Gaussian vector
$\mathbf{x} \in {\mathbb{C}^{N \times 1}}$ with zero mean and covariance matrix ${\bf{R}}_N$.
$\rm{null}{\left(\mathbf{A}\right)}$ denotes the null space of matrix $\mathbf{A}$.

\section{System Model}
We study the MIMOME wiretap channel with a multiple-antenna transmitter (Alice), a multiple-antenna
 intended receiver (Bob), and a multiple-antenna  eavesdropper (Eve), where the corresponding numbers of antennas are denoted by
 $N_t$, $N_r$, and $N_e$, respectively. The signals
 received at Bob and Eve are denoted by ${\mathbf{y}}_b$ and ${\mathbf{y}}_e$, respectively,
and can be written as
\begin{equation}\label{main}
{\mathbf{y}}_b  =  {\mathbf{H}}_{ba} {\mathbf{Gx}}_a  + {\mathbf{n}}_b
\end{equation}
\begin{equation}\label{eave}
{\mathbf{y}}_e  =  {\mathbf{H}}_{ea} {\mathbf{Gx}}_a  + {\mathbf{n}}_e
\end{equation}
where ${\mathbf{x}}_a = [x_1, x_2, \cdots,x_{N_t}]^T \in \mathbb{C}^{N_t \times 1}$ denotes
the transmitted signal vector having zero mean and the identity matrix as
covariance matrix, and ${\mathbf{H}}_{ba} \in \mathbb{C}^{N_r \times N_t}$ and ${\mathbf{H}}_{ea} \in \mathbb{C}^{N_e \times N_t}$
denote the channel matrices between Alice and Bob and between Alice and Eve, respectively.
 The complex independent identically distributed
(i.i.d.) vectors ${\mathbf{n}}_b \sim {\cal CN}(0, \; \sigma_b^2 {\bf{I}}_{N_r})$
and ${\mathbf{n}}_e \sim {\cal CN}(0, \; \sigma_e^2 {\bf{I}}_{N_e})$
represent the channel noises at Bob and Eve, respectively.
${\mathbf{G}} \in \mathbb{C}^{N_t \times N_t}$ is
a linear precoding matrix that has to be optimized
for maximization of the secrecy rate.
The precoding matrix has to satisfy
the power constraint
\begin{equation}\label{power_constraint}
{\tr}\left\{ {E\left[ {{\mathbf{Gx}}_a\mathbf{x}_a^H  {\mathbf{G}}^H  } \right]} \right\} = {\tr}\left\{ {{\mathbf{GG}}^H  } \right\} \le P.
\end{equation}

In order to be able to present the basic idea
behind PG-GSVD design in the simplest manner possible,
we assume in this paper that the perfect
instantaneous CSI of both the intended receiver and the eavesdropper are available at the transmitter
in this paper. This assumption applies for the case where
the transmitter intends to send a private message to a particular user in the system while regarding another user as eavesdropper, i.e.,
the eavesdropper is an idle user of the system \cite{Gursoy2012TCom,Zeng2016TWC}.
Based on the deterministic equivalent channel model
for the large system limit derived in \cite{Wu2015TWC},
the PG-GSVD design in this paper can be easily extended
to the case  where only the statistical CSI of the eavesdropper is available at the transmitter.

When the transmitter has perfect
instantaneous  knowledge of the eavesdropper's channel, the achievable secrecy rate is given by  \cite{Khisti2010TIT_2}
\begin{equation}\label{secrecy_capacity}
C_{\rm{sec} } =  {\mathop {\max }\limits_{\tr\left( {{\bf{GG}}^H } \right) \le P } }  R_{\rm sec}({\mathbf{G}})
\end{equation}
\begin{equation}\label{rate}
R_{\rm sec}({\mathbf{G}}) = I \left( {{\mathbf{y}}_b ;{\mathbf{x}}_a} \right) - I \left( {{\mathbf{y}}_e ;{\mathbf{x}}_a} \right)
\end{equation}
where $I(\mathbf{y}; \mathbf{x})$ denotes the mutual information  between input $\mathbf{x}$ and output $\mathbf{y}$.

The goal of this paper is to optimize the transmit precoding matrix $\mathbf{G}$  for maximization of
 the secrecy rate in (\ref{rate}) when the transmit symbols ${\mathbf{x}}_a $ are drawn from a discrete constellation set with $M$ equiprobable points
such as $M$-ary quadrature amplitude modulation (QAM) and $N_t$ is large.

\section{Low Complexity Precoder Design with Instantaneous CSI of the Eavesdropper}
In this section, we first provide some useful definitions which will be used in the
subsequent analysis. Then, we analyze the rate loss
of the GSVD design \cite{Bashar2012TCom} compared to the maximal rate for finite alphabet inputs in the high SNR regime.
Finally, we propose a PG-GSVD precoder to compensate this performance loss with low complexity.

\subsection{Some Useful Definitions} \label{sec:definition}
Let us introduce some useful definitions for the subsequent analysis.
\begin{definition}
Similar to \cite{Bashar2012TCom,Khisti2010TIT_2}, we define the following subspaces
\begin{eqnarray*}
\mathcal{S}_{ba}    & = {\rm null}\left(\mathbf{H}_{ba}\right)^{\bot}&\cap~\,\,{\rm null}\left(\mathbf{H}_{ea}\right)  \\
\mathcal{S}_{be}  &={\rm null}\left(\mathbf{H}_{ba}\right)^{\bot}&\cap~\,\,{\rm null}\left(\mathbf{H}_{ea}\right)^{\bot}\\
\mathcal{S}_{ea}   &= {\rm null}\left(\mathbf{H}_{ba}\right)&\cap~\,\,{\rm null}\left(\mathbf{H}_{ea}\right)^{\bot}\\
\mathcal{S}_{n}   &={\rm null}\left(\mathbf{H}_{ba}\right)&\cap~\,\,{\rm null}\left(\mathbf{H}_{ea}\right).
\end{eqnarray*}

\end{definition}

\noindent Define $k={\rm rank}\left(\left[\begin{array}{cc}
\mathbf{H}_{ba}^{H} & \mathbf{H}_{ea}^{H}\end{array}\right]^{H}\right)$ and hence ${\rm dim}\left(\mathcal{S}_{n}\right)= N_t -k$.
In addition, define $r={\rm dim}\left(\mathcal{S}_{ba}\right)$ and $s={\rm dim}\left(\mathcal{S}_{be}\right)$.
Therefore, ${\rm dim}\left(\mathcal{S}_{ea}\right)=k-r-s$.

\begin{definition}
Following \cite{Khisti2010TIT_2}, we define  the GSVD of the pair $\left(\mathbf{H}_{ba}, \mathbf{H}_{ea} \right)$
as follows:
\begin{equation} \label{eq:sigma_ba}
\mathbf{H}_{ba}= {\mathbf{U}}_{ba}~\boldsymbol{\Sigma}_{ba}
\kbordermatrix {~ & k & N_t -k \cr
		               ~ &\boldsymbol{\Omega}^{-1}  &  \mathbf{0} \cr}
~{\mathbf{U}}_{a}^{H}
\end{equation}
\begin{equation} \label{eq:sigma_ea}
\mathbf{H}_{ea}= {\mathbf{U}}_{ea}~\boldsymbol{\Sigma}_{ea}
\kbordermatrix {~ & k & N_t -k \cr
		               ~ &\boldsymbol{\Omega}^{-1}  &  \mathbf{0} \cr}
~{\mathbf{U}}_{a}^{H}
\end{equation}
where ${\mathbf{U}}_{a} \in \mathbb{C}^{N_t \times N_t}$, ${\mathbf{U}_{ba}} \in \mathbb{C}^{N_r \times N_r}$,
and ${\mathbf{U}}_{ea}\in \mathbb{C}^{N_e \times N_e}$ are unitary matrices.
$\boldsymbol{\Omega} \in \mathbb{C}^{k\times k}$ is a non-singular matrix with diagonal elements $\omega_i$, $i=1,\ldots,k$.
 $\boldsymbol{\Sigma}_{ba}\in\mathbb{C}^{N_r \times k}$ and $\boldsymbol{\Sigma}_{ea}\in \mathbb{C}^{N_e\times k}$ can be expressed as

\begin{equation}
\boldsymbol{\Sigma}_{ba} = \kbordermatrix {~		   & k-r-s	 & s 		& r 		\cr
										N_r - r - s & \mathbf{0} 	& \mathbf{0}  		& \mathbf{0} 	\cr
										s                 & \mathbf{0}  	& \mathbf{D}_b 	& \mathbf{0} 	\cr
										r                 & \mathbf{0}  	& \mathbf{0}      	& \mathbf{I}_r 	\cr}
\end{equation}

\begin{equation}
\boldsymbol{\Sigma}_{ea} = \kbordermatrix{~		   & k-r-s	 & s 		& r 		\cr
										k - r - s      & \mathbf{I}_{k-r-s} 	& \mathbf{0} 		& \mathbf{0} 	\cr
										s                 & \mathbf{0} 	& \mathbf{D}_e 	& \mathbf{0} 	\cr
										N_e-k+r    & \mathbf{0} 	& \mathbf{0}     	& \mathbf{0}	\cr}
\end{equation}

\end{definition}

\noindent where $\mathbf{D}_b={\rm diag}\left(\left[b_1,\ldots,b_s\right]\right) \in \mathbb{R}^{s \times s} $ and
$\mathbf{D}_e={\rm diag}\left(\left[e_1,\ldots,e_s\right]\right) \in \mathbb{R}^{s \times s}$
are diagonal matrices with real valued entries. The diagonal elements of $\mathbf{D}_b$ and $\mathbf{D}_e$ are ordered
as follows:
\[0<b_1\leq b_2 \leq \ldots \leq b_s < 1\] \[1>e_1\geq e_2 \geq \ldots \geq e_s > 0\] and \[b_p^2 + e_p^2 =1,~\mbox{for}~p=1,\ldots,s .\]

\subsection{Performance Loss of the GSVD Design}
The precoding matrix for the GSVD design can be expressed as \cite{Bashar2012TCom}
\begin{equation}
\mathbf{G} = \mathbf{U}_{a} \mathbf{A} \mathbf P^{\frac{1}{2}}
\label{eq:precoding_matrix}
\end{equation}
where  $\mathbf{P} = {\rm diag} \left(p_1,\ldots,p_{N_t}\right)$ represents
a diagonal power allocation matrix and $\mathbf{A}$ is given by
\begin{equation}
\mathbf{A} = \kbordermatrix{
~ 		& k 							& N_t - k 	\cr
k		& \boldsymbol{\Omega}		& \mathbf{0}		\cr
N_t -k	& \mathbf{0}						& \mathbf{0}		\cr
}.
\end{equation}
For the GSVD precoder design in (\ref{eq:precoding_matrix}), the
received signals ${\mathbf{y}}_b$ and ${\mathbf{y}}_e$ in (\ref{main}) and (\ref{eave})
can be re-expressed as
\begin{eqnarray}
\mathbf{\tilde{y}}_{b} &=&  ~\boldsymbol{\Sigma}_{ba}
\kbordermatrix {~ & k     	&   N_t -k 	\cr
		                ~ & \mathbf{I}_k 	&  \mathbf{0}		\cr}
~\mathbf{P}^{\frac{1}{2}}~\mathbf{x}_a + \mathbf{\tilde{n}}_{b} \label{eq:bob_precoded}\\
\mathbf{\tilde{y}}_{e} &=& ~\boldsymbol{\Sigma}_{ea}
\kbordermatrix {~ & k     	&  N_t -k 	\cr
		                ~ & \mathbf{I}_k 	&   \mathbf{0} 		\cr}
~\mathbf{P}^{\frac{1}{2}}~ \mathbf{x}_a +\mathbf{\tilde{n}}_{e}  \label{eq:eve_precoded}
\end{eqnarray}
where $\tilde{\mathbf y}_{b} = {\mathbf{U}}_{ba}^H \mathbf{y}_b$, $\tilde{\mathbf y}_{e} ={\mathbf{U}}_{ea}^H \mathbf{y}_e$,
 $\tilde{\mathbf{n}}_{b} = {\mathbf{U}}_{ba}^H \mathbf{n}_b$, and $\tilde{\mathbf n}_{e} = {\mathbf{U}}_{ea}^H \mathbf{n}_e$.

Define $N_1 = {\rm rank}\left( {\mathbf{H}}_{ba} \right)$ and $N_2 = {\rm rank}\left( {\mathbf{H}}_{ea} \right)$.
In the following theorem, we analyze the performance of the GSVD  design for finite alphabet inputs in the
high SNR regime.
\begin{theorem}\label{GSVD_loss}
In the high SNR regime  ($P \rightarrow \infty$), for the GSVD design in (\ref{eq:precoding_matrix}),
the achievable secrecy rate $R_{\rm sec, high}$ for finite alphabet signals
is upper bounded by
\begin{equation}\label{rate_high}
R_{\rm sec, high} \leq  N_1 \log_2 M  \ \textrm{b/s/Hz}.
\end{equation}
\begin{proof}
See Appendix \ref{GSVD_loss_proof}.
\end{proof}
\end{theorem}

Theorem \ref{GSVD_loss} indicates that the GSVD design may result in
a severe performance loss for finite alphabet inputs in the high SNR regime.
For example, if $N_t > N_r$, which is a typical scenario for large-scale MIMO systems \cite{Marzetta2010TWC,Larsson2014CM},
the GSVD design will cause a rate loss of at least $(N_t -N_r) \log_2 M$ b/s/Hz compared to
the maximal rate in the high SNR regime. The precoder design in \cite{Wu2012TVT} avoids this performance loss by directly optimizing
the precoder matrix $\mathbf{G}$.  However, this results in an intractable implementation complexity for large-scale MIMO
systems. Inspired by the idea of decoupling and grouping of point-to-point MIMO channels
for finite alphabet inputs \cite{Mohammed2011TIT,Ketseoglou2015TWC,Wu2016ICC}, we propose a PG-GSVD
precoder design to prevent  the performance loss of the GSVD design
while retaining a low complexity  in large-scale MIMOME channels.

\subsection{PG-GSVD Precoder Design} \label{sec:PG-GSVD}
As indicated in \cite{Wu2016ICC}, in order to decouple the MIMO channels
into $N_t$ parallel subchannels, the MIMO channel matrix has to be an $N_t \times N_t$ matrix.
However,  $\boldsymbol{\Sigma}_{ba}$ and $\boldsymbol{\Sigma}_{ea}$ in (\ref{eq:bob_precoded}) and (\ref{eq:eve_precoded}) are
$N_r \times N_t$ and $N_e \times N_t$ matrices, respectively. As a result,
we need to add to or remove from $\mathbf{\tilde{y}}_{b}$, $\boldsymbol{\Sigma}_{ba}$,
$\mathbf{\tilde{y}}_{e}$, and $\boldsymbol{\Sigma}_{ea}$ some zeros in (\ref{eq:bob_precoded}) and (\ref{eq:eve_precoded}).
To this end, we define
\begin{align}
\mathbf{\hat{y}}_{b} = \kbordermatrix{
~ & ~ \cr
k - r -s  & \mathbf{0}  \cr
r + s & \tilde{\tilde{\mathbf{y}}}_{b}^H  \cr
 N_t - k & \mathbf{0}},
\end{align}
where $\tilde{\tilde{\mathbf{y}}}_{b} \in \mathbb{C}^{(r + s) \times 1}$
is composed of the last $r + s$ elements of $\mathbf{\tilde{y}}_{b}$.
Furthermore, we define $\boldsymbol{\omega} = \left[ \omega_1,\cdots,\omega_k \ \mathbf{0}^{T} \right]^H \in \mathbb{C}^{N_t \times 1}$,
$\mathbf{\hat{y}}_{e} = \left[\tilde{\mathbf y}_{e}^H \ \mathbf{0}^{T} \right]^H \in \mathbb{C}^{N_t \times 1}$,
$\mathbf{\hat{n}}_{b} \sim {\cal CN}(0, \; \sigma_b^2 {\bf{I}}_{N_t})$, and $\mathbf{\hat{n}}_{e}  \sim {\cal CN}(0, \; \sigma_e^2 {\bf{I}}_{N_t}) $.
Define two diagonal matrices
\begin{equation}\label{eq:sigma_ba_hat}
\boldsymbol{\hat{\Sigma}}_{ba} = \kbordermatrix {~		   &  k -r-s	 & s 		& r 	& N_t - k	\cr
										k - r - s & \mathbf{0} 	& \mathbf{0}  		& \mathbf{0} & \mathbf{0}	\cr
										s                 & \mathbf{0}  	& \mathbf{\hat{D}}_b 	& \mathbf{0} & \mathbf{0}	\cr
										r                 & \mathbf{0}  	& \mathbf{0}      	& \mathbf{R}_r 	& \mathbf{0}  \cr
                                       N_t -k                 & \mathbf{0}  	& \mathbf{0}      	& \mathbf{0} 	& \mathbf{0}  \cr  }
\end{equation}

\begin{equation} \label{eq:sigma_ea_hat}
\boldsymbol{\hat{\Sigma}}_{ea} = \kbordermatrix{~		   & k-r-s	 & s 		& N_t - k + r		\cr
										k - r - s      & \mathbf{R}_{k-r-s} 	& \mathbf{0} 		& \mathbf{0}  	\cr
										s                 & \mathbf{0} 	& \mathbf{\hat{D}}_e 	& \mathbf{0} 	\cr
										N_t -k+r    & \mathbf{0} 	& \mathbf{0}     	& \mathbf{0} 	\cr
                                         }
\end{equation}
where the elements of $\mathbf{\hat{D}}_b$,  $\mathbf{R}_r$,
$\mathbf{R}_{k-r-s}$, and $\mathbf{\hat{D}}_e$
are obtained from the following two equations
\begin{align}
& \hspace{-1.5cm} \left[\boldsymbol{\hat{\Sigma}}_{ba}\right]_{(k-r-s+i)(k-r-s+i)} =  \left[\boldsymbol{{\Sigma}}_{ba}\right]_{(N_r -r-s + i)(k-r-s+i)}/\sqrt{\omega_i}, \nonumber \\
& \hspace{1cm} i = 1,\cdots,s+r  \label{eq:sigma_ba_hat_ele} \\
\left[\boldsymbol{\hat{\Sigma}}_{ea}\right]_{ii} = &  \left[\boldsymbol{{\Sigma}}_{ea}\right]_{ii}/ \sqrt{\omega_i}, \  i = 1,\cdots,k-r. \label{eq:sigma_ea_hat_ele}
\end{align}
We divide the transmit signal $\mathbf{x}_a$ into $S$ streams
and let $N_{s}=N_{t}/S$\footnote{For convenience, we assume $N_{s}=N_{t}/S$ is an integer in this paper. If
$N_{t}/S$ is not an integer, we can easily obtain an integer $N_s$ by adding zeros in (\ref{eq:sigma_ba_hat})
and (\ref{eq:sigma_ea_hat}).}.
We define the set $\left\{\ell_{1},\ldots,\ell_{N_{t}}\right\}$
as a permutation of $\left\{1,\ldots,N_{t}\right\}$.
${\bf{P}}_s \in \mathbb{C}^{N_{s} \times N_{s}}$
and $\mathbf{V}_s \in \mathbb{C}^{N_{s} \times N_{s}}$,
$s = 1,\ldots,S$, denote a diagonal
and a unitary matrix, respectively.
$\mathbf{V} \in \mathbb{C}^{N_t \times N_t}$ denotes a unitary matrix.
For the proposed PG-GSVD precoder, we set $\mathbf{G}$ as follows
\begin{equation}
\mathbf{G} = \mathbf{U}_{a} \mathbf{A} \mathbf P^{\frac{1}{2}} \mathbf{V}.
\label{eq:precoding_matrix_gsvd}
\end{equation}

We set
\begin{align}\label{eq:P_pair}
\left[\boldsymbol{\omega}\right]_{\ell_{j}\ell_{j}}  {\left[ {\bf{P}} \right]_{\ell_{j}\ell_{j}} } =  \left[ {{{\bf{P}}}}_s \right]_{ii},
\end{align}
where $i = 1,\ldots,N_{\mathrm s}$, $s=1,\ldots,S$, and $j = (s - 1) N_{\mathrm s} + i$.
Based on (\ref{eq:precoding_matrix_gsvd}) and (\ref{eq:P_pair}), the power constraint in (\ref{power_constraint}) is
equivalent to $\sum \nolimits_{s = 1}^{S} \tr \left({\mathbf{P}_s} \right) \leq P$.

Also, we set
\begin{align}\label{eq:V_pair}
& \!\!\!\!  \left[ {{{\bf{V}}}} \right]_{\ell_i \ell_j} =  \\
& \left\{ \begin{array}{l}
   {\left[ {{{\bf{V}}_{s}}} \right]_{mn}} \quad {\rm if} \ i = (s - 1) N_{\mathrm s} + m , \ j = (s - 1) N_{\mathrm s}  + n   \\
0 \qquad \quad\;\; {\rm otherwise}
 \end{array} \right. \nonumber
\end{align}

 \begin{alg} \label{Gradient_Pair}
      Maximizing $R_{\rm sec}({\mathbf{G}})$  with respect to $\mathbf{P}_s$ and $\mathbf{V}_s$.

    \vspace*{1.5mm} \hrule \vspace*{1mm}
      \begin{enumerate}

    \itemsep=0pt

    \item Initialize $\mathbf{P}_s$ with $ \sum\nolimits_{s = 1}^S \tr\left(\mathbf{P}_{{s}}^{(0)}\right) = N_t$ and $\mathbf{V}_{{s}}^{(0)}$ for  $s = 1,\ldots,S$.
    Set $N_{\rm iter}$ and $\varepsilon$ as the maximum iteration number and a threshold, respectively.

    \item  Initialize $R_{\rm sec}({\mathbf{G}})^{(1)}$ based on (\ref{eq:I_pair}).  Set counter $n = 1$.

    \item Update ${\mathbf{P}}_{{s}}^{(n)}$ for $s= 1,\ldots,S$ along
    the gradient decent direction ${\nabla _{{\bf{P}}_{s}}} R({\mathbf{G}})$.

    \item Normalize $ \sum\nolimits_{s=1}^{S} \tr\left({\mathbf{P}}_{{s}}^{(n)}\right) = P$.

    \item Update $\mathbf{V}_{{s}}^{(n)}$ for $s= 1,\ldots,S$ along the gradient descent direction ${\nabla _{{\bf{V}}_{s}}} R({\mathbf{G}})$.

    \item Compute $R_{\rm sec}({\mathbf{G}})^{(n+1)}$ based on (\ref{eq:I_pair}).
    If $R_{\rm sec}({\mathbf{G}})^{(n+1)}  - R_{\rm sec}({\mathbf{G}})^{(n)} > \varepsilon$
    and $n \leq N_{\rm iter}$,  set $n = n + 1$ and repeat Steps $3$--$5$;

     \item Compute $\mathbf{P}$ and $\mathbf{V}$ based on (\ref{eq:P_pair}) and (\ref{eq:V_pair}).
     Set $\qG = \mathbf{U}_{a} \mathbf{A} \mathbf P^{\frac{1}{2}} \mathbf{V}$.

    \vspace*{1mm} \hrule

     \end{enumerate}

   \end{alg}

    \null
    \par

\noindent where $m = 1,\ldots, N_{s}$, $n = 1,\ldots,N_{s}$, $s = 1,\ldots, S$,
$i =  1,\ldots, N_{t}$, and $j = 1,\ldots, N_{t}$.
Finally, we let
\begin{align}\label{eq:x_pair}
\left[\mathbf{x}_{s}\right]_{i} =  \left[ \mathbf{x}_a \right]_{\ell_{j}} .
\end{align}
Based on (\ref{eq:precoding_matrix_gsvd})--(\ref{eq:x_pair}) and a paring scheme $\left\{\ell_{1},\ldots,\ell_{N_{t}}\right\}$,
the equivalent received signals
at Bob and Eve can be decoupled as follows
  \begin{eqnarray}
 \left[ \mathbf{\hat{y}}_{b} \right]_{\ell_j} & =  \left[\boldsymbol{\hat\Sigma}_{ba} \right]_{\ell_j\ell_j}  \left[\mathbf{\hat{x}}\right]_{\ell_j}
 + \left[\mathbf{\hat{n}}_{b} \right]_{\ell_j}   \label{eq:bob_precoded_low}   \\
  \left[ \mathbf{\hat{y}}_{e} \right]_{\ell_j} & =  \left[\boldsymbol{\hat\Sigma}_{ea} \right]_{\ell_j\ell_j}  \left[\mathbf{\hat{x}}\right]_{\ell_j}
  + \left[\mathbf{\hat{n}}_{e} \right]_{\ell_j}
   \label{eq:eve_precoded_low}
  \end{eqnarray}
 where
 \begin{align}\label{eq:x_hat}
   \left[\mathbf{\hat{x}}\right]_{\ell_j} = \left[ \mathbf{P}_s^{\frac{1}{2}} \mathbf{V}_s \mathbf{x}_s \right]_i
 \end{align}
 for $i = 1,\ldots,N_{s}$, $s = 1,\ldots, S$, and  $j = (s - 1) N_{s} + i$.
From (\ref{eq:bob_precoded_low})
 and (\ref{eq:eve_precoded_low}), we observe that the transmit signal has been
divided into $S$ independent
 groups. In each group, the equivalent signal dimension is $N_s \times 1$.
 We further define $ \left[\mathbf{\hat{y}}_b \right]_{\ell_j} = \left[\mathbf{y}_{b,s}\right]_i$ and
 $ \left[\mathbf{\hat{y}}_e \right]_{\ell_j} = \left[\mathbf{y}_{e,s}\right]_i$.

Based on (\ref{eq:bob_precoded_low}) and (\ref{eq:eve_precoded_low}), the secrecy rate in (\ref{rate}) can be expressed as
\begin{equation}\label{eq:I_pair}
  R_{\rm sec}({\mathbf{G}}) = \sum\limits_{s = 1}^S \left( I\left( {{\bf{y}}_{b,s}};\mathbf{x}_{s} \right) - I\left({{\bf{y}}_{e,s}};\mathbf{x}_{s} \right) \right).
\end{equation}
The gradients of $I\left( {{\bf{y}}_{b,s}};\mathbf{x}_{s} \right)$ and $I\left({{\bf{y}}_{e,s}};\mathbf{x}_{s} \right)$
with respect to $\mathbf{P}_{s}$ and $\mathbf{V}_{s}$ can be found in
 \cite[Eq. (22)]{Palomar2006TIT}, based on  which an iterative algorithm  can be derived
for maximizing  $ R_{\rm sec}({\mathbf{G}})$, as given in Algorithm 1.

For the precoder design with finite alphabet inputs, the computational complexity
is mainly dominated by the required number of additions in calculating the mutual information and
the MSE matrix when $N_t$ is large \cite{Wu2015TWC}. Considering the decoupled structure in (\ref{eq:I_pair}),
the computational complexity of Algorithm \ref{Gradient_Pair} grows
linearly with $ S M^{2 N_{s}}$.  However, the computational complexity of
the algorithm in \cite{Wu2012TVT} scales linearly with  $ M^{2 N_{t}}$.
We observe that for large-scale MIMO systems when $N_t$ is large, the computational complexity
of Algorithm \ref{Gradient_Pair} is by orders of magnitude lower than the algorithm in \cite{Wu2012TVT}.

For the PG-GSVD design in (\ref{eq:precoding_matrix_gsvd}), we have the following theorem.

\begin{theorem}\label{PG-GSVD_high}
If the inequality $ (k - N_2) N_s \geq N_t$ holds,
then we can always find a permutation $\left\{\ell_{1},\ldots,\ell_{N_{t}}\right\}$  for the PG-GSVD
design in (\ref{eq:precoding_matrix_gsvd}), which achieves  $R_{\rm sec, high} =  N_t \log_2 M $ b/s/Hz
in the high SNR regime.
\begin{proof}
See Appendix \ref{Proof_PG-GSVD_high}.
\end{proof}
\end{theorem}

The algorithm in \cite{Wu2012TVT} is equivalent to setting $N_s = N_t$ in Algorithm \ref{Gradient_Pair}. Therefore, as long as $k - N_2 \neq 0$, it can
compensate the performance loss of the GSVD design and achieve the saturation rate  $N_t \log_2 M$ b/s/Hz in the high SNR regime,
as shown in \cite[Figs. 1, 2]{Wu2012TVT}. However, in this case, the computational complexity of the algorithm in \cite{Wu2012TVT} grows
exponentially with $N_t$.  This is prohibitive in large-scale MIMO systems.
For typical large-scale MIMO systems, we have $N_t > N_2$ \cite{Marzetta2010TWC,Larsson2014CM}, which implies $k - N_2 \neq 0$.
As a result, by properly choosing $N_s$, we can reach a favorable trade-off between complexity and secrecy rate
performance.

\section{Numerical Results}
We set $\sigma_b = \sigma_e$ and define ${\rm SNR} =  P/(N_r \sigma_b^2)$. Furthermore, we use $N_t \times N_r \times N_e$
to denote the simulated wiretap channel.

\subsection{Scenarios with Instantaneous CSI of the Eavesdropper}
\begin{figure}[!t]
\centering
\includegraphics[width=0.4\textwidth]{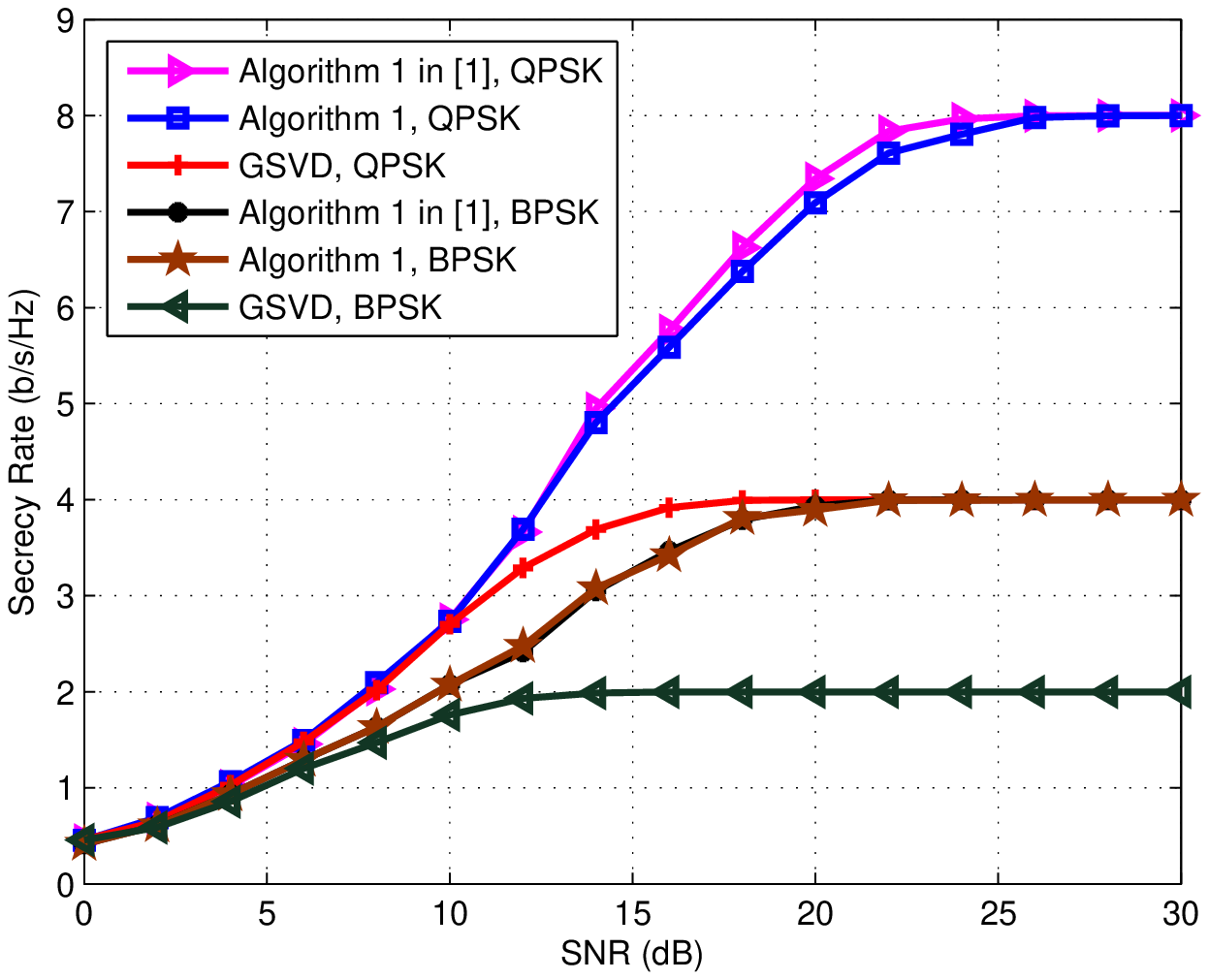}
\vspace{-10pt}
 \captionstyle{flushleft}
 \caption{Secrecy rate versus SNR for the $4\times3\times2$ wiretap channel for different precoder designs and different modulation schemes  for $N_{s} = 2$.}
\label{wiretap_421}
\end{figure}

\begin{figure}[!t]
\centering
\includegraphics[width=0.4\textwidth]{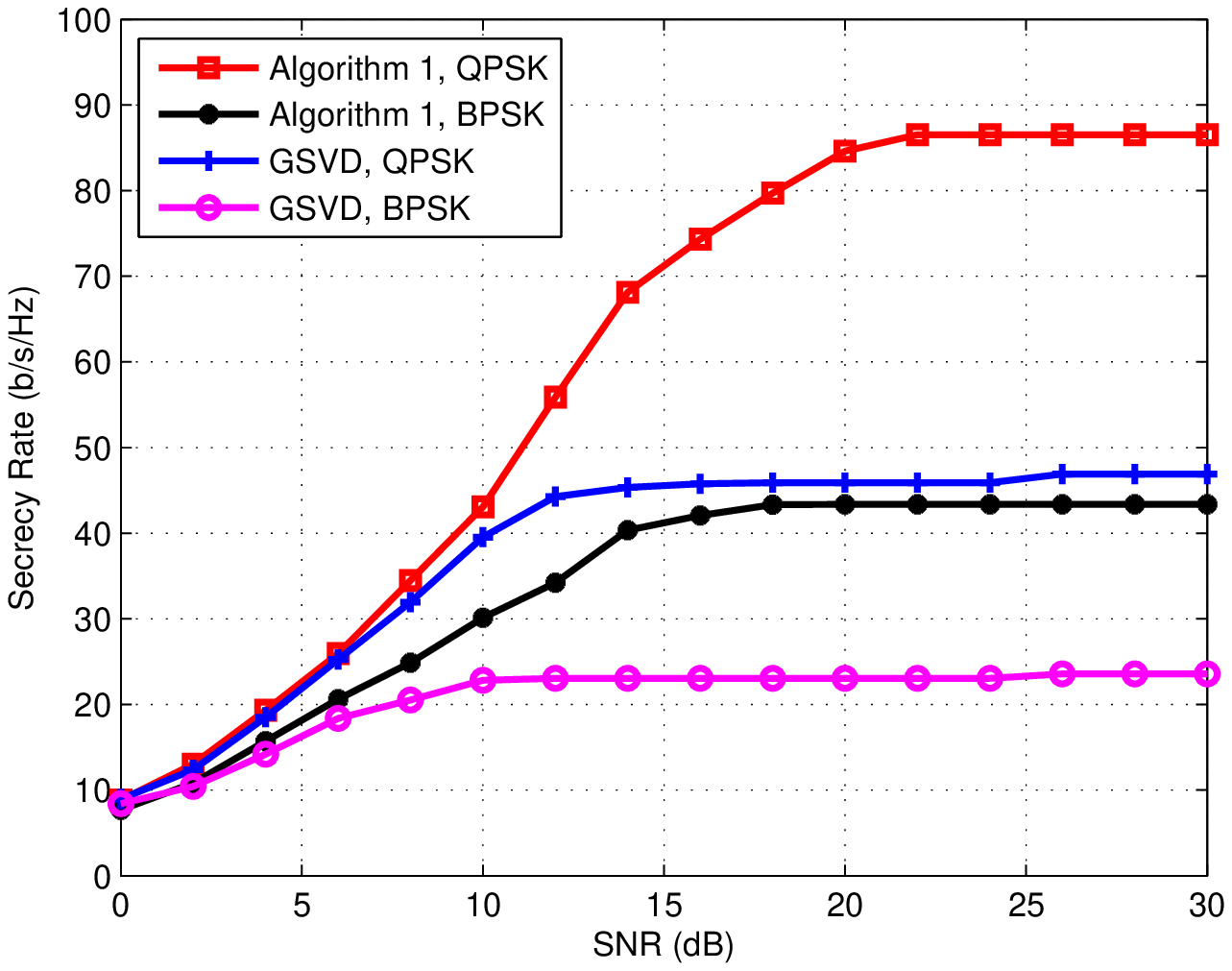}
 \vspace{-10pt}
 \captionstyle{flushleft}
\caption{Secrecy rate versus SNR for the $64\times48\times48$ wiretap channel for different precoder designs and different modulation schemes  for $N_{s} = 2$.}
\label{wiretap_64_48_48}
\end{figure}

In this subsection, the elements of $\mathbf{H}_{ba}$
and $\mathbf{H}_{ea}$ are generated independently and randomly.
Tables I and II compare the computational complexities of the different schemes
for the systems considered in Figures \ref{wiretap_421} and \ref{wiretap_64_48_48}, respectively.

\begin{table}[!t]
\centering
\caption{Number of additions required for calculating the mutual information and the MSE matrix for the system considered in Figure \ref{wiretap_421}.}
\vspace*{1.5mm}
\begin{tabular}{|c|c|c|}
\hline
$4\times3\times2$   & BPSK &  QPSK     \\ \hline
 GSVD &  8    &   16 \\ \hline
Algorithm 1 &  32   &   512 \\  \hline
 Algorithm 1 in \cite{Wu2012TVT} &   256   &   65536     \\ \hline
\end{tabular}
\end{table}

\begin{table}[!t]
\centering
\caption{Number of additions required for calculating the mutual information and the MSE matrix for the system considered in Figure \ref{wiretap_64_48_48}.}
\vspace*{1.5mm}
\begin{tabular}{|c|c|c|}
\hline
$64\times48\times48$   & BPSK &  QPSK     \\ \hline
 GSVD &  128    &   256 \\ \hline
Algorithm 1 &  512   &   8192 \\  \hline
 Algorithm 1 in \cite{Wu2012TVT} &  3.04e+038     &  1.15e+077    \\ \hline
\end{tabular}
\end{table}

Figure \ref{wiretap_421} plots the secrecy rate for the $4\times3\times2$ wiretap channel
for different precoder designs and different modulation schemes for $N_s =2$.
 We observe from Figure \ref{wiretap_421} that Algorithm 1
achieves a similar performance as the precoder design in \cite{Wu2012TVT} but with orders of magnitude lower computational complexity
as indicated in Table I.
Both designs achieve the maximal rate $ N_t \log_2 M$ b/s/Hz in the high SNR regime as indicated by Theorem \ref{PG-GSVD_high}.
In contrast, the GSVD design yields an obvious rate loss in the high SNR regime.
For the channels of Bob and Eve, we have $D_{b,1} = 0.57$, $D_{e,1} = 0.81$.
As explained  in Example 1, the GSVD design sets $p_1 = p_2 =0$ in this case.
Therefore, the GSVD design suffers from a $2 \log_2 M$ b/s/Hz rate loss in the high SNR regime as shown in Figure  \ref{wiretap_421}.

In Figure \ref{wiretap_64_48_48}, we show the secrecy rate for the $64\times48\times48$
wiretap channel
for different precoder designs and different modulation schemes for $N_s =2$.
As indicated in Table II,
the computational complexity of the precoder design in \cite{Wu2012TVT} is prohibitive in this case and no results can
be shown. We observe that
the secrecy rate of the GSVD design is lower than the upper bound given in Theorem \ref{GSVD_loss}.
This is because for the GSVD design, as indicated in \cite[Eq. (12)]{Bashar2012TCom}, only the non-zero subchannels of Bob which are
stronger than the corresponding subchannels
of Eve can be used for transmission.  The $b_i$, $i = 1,\ldots,s$, in (\ref{eq:sigma_ba}) are in ascending order while the $e_i$, $i = 1,\ldots,s$,
in (\ref{eq:sigma_ea}) are in descending order.
Therefore, a large proportion of Bob's non-zero subchannels may be abandoned by the GSVD design for large-scale MIMO channels.
As a result, Algorithm 1 achieves significantly higher secrecy rates than the GSVD design.

\section{Conclusion}
In this paper, we have investigated the linear precoder design
for large-scale MIMOME wiretap channels with finite alphabet signals.
We derived an upper bound on the secrecy rate
for the GSVD design in the high SNR regime. The derived expression reveals that
the GSVD design may lead to a serious performance loss. Based on this, we proposed a PG-GSVD design to overcome
the negative properties of the GSVD design while retaining an affordable computational
complexity for large-scale MIMO systems.
Simulations indicated that the proposed design performs well in large-scale MIMOME
wiretap channels and achieves substantial secrecy rate gains
compared to the GSVD design for finite alphabet inputs.

\appendices

\section{Proof of Theorem \ref{GSVD_loss}} \label{GSVD_loss_proof}
Based on (\ref{eq:sigma_ba}) and (\ref{eq:bob_precoded}), $I \left( {{\mathbf{y}}_b ;{\mathbf{x}}_a} \right)$ in (\ref{rate}) for the GSVD design becomes
\begin{equation} \label{rate_b_GSVD}
I\left( {{{\bf{y}}_b};{{\bf{x}}_a}} \right) = \sum\limits_{i = 1}^s {I\left( {b_i^2{p_{k - r - s + i}}} \right)}  + \sum\limits_{i = 1}^r {I\left( {{p_{k - s + i}}} \right)}
\end{equation}
where $I(\gamma) =  I(x; \sqrt{\gamma} x + n)$. Therefore, for $P \rightarrow \infty$, we obtain
\begin{equation} \label{rate_b_GSVD}
\mathop {\lim }\limits_{P \to \infty }  I\left( {{{\bf{y}}_b};{{\bf{x}}_a}} \right) \leq (s + r) \log_2 (M).
\end{equation}
According to  Inclusion--Exclusion Principle \cite{Andrews1971book}, we know
\begin{align} \label{Dim_set}
{\rm dim}\left(\mathcal{S}_{ba}\right) +  {\rm dim}\left(\mathcal{S}_{be}\right) = {\rm dim} \left(\mathcal{S}_{ba} \cup \mathcal{S}_{be}  \right) -
{\rm dim} \left(\mathcal{S}_{ba} \cap \mathcal{S}_{be}\right).
\end{align}
For the subspaces $\mathcal{S}_{ba}$ and $\mathcal{S}_{be}$, we have
\begin{subequations} \label{S_set_total}
\begin{equation*} \label{S_set_no}
\hspace{-5cm} \mathcal{S}_{ba} \cap \mathcal{S}_{be}  =
\end{equation*}
\begin{equation} \label{S_set}
 \left({\rm null}\left(\mathbf{H}_{ba}\right)^{\bot}   \cap  {\rm null}\left(\mathbf{H}_{ea}\right)\right) \cap
\left( {\rm null}\left(\mathbf{H}_{ba}\right)^{\bot} \cap {\rm null}\left(\mathbf{H}_{ea}\right)^{\bot}\right)
\end{equation}
\begin{equation}\label{eq:Set_2}
\hspace{-0.1cm} = \left({\rm null}\left(\mathbf{H}_{ba}\right)^{\bot}   \cap  {\rm null}\left(\mathbf{H}_{ea}\right)\right)  \cap
\left(  {\rm null}\left(\mathbf{H}_{ea}\right)^{\bot} \cap{\rm null}\left(\mathbf{H}_{ba}\right)^{\bot} \right)
\end{equation}
\begin{equation}\label{eq:Set_3}
\hspace{-0.1cm} =  {\rm null}\left(\mathbf{H}_{ba}\right)^{\bot}  \cap  \left(\left( {\rm null}\left(\mathbf{H}_{ea}\right)\right) \cap  {\rm null}\left(\mathbf{H}_{ea}\right)^{\bot}    \right)  \cap {\rm null}\left(\mathbf{H}_{ba}\right)^{\bot}
\end{equation}
\begin{equation}\label{eq:Set_4}
\hspace{-8.4cm}  = \varnothing
\end{equation}
\end{subequations}
where (\ref{eq:Set_2}) and (\ref{eq:Set_3}) are obtained based on the properties of intersections \cite{Halmos1960book}.

Also, we have
\begin{subequations} \label{S_set_cup}
\begin{equation*}
\hspace{-5cm} \mathcal{S}_{ba} \cup \mathcal{S}_{be}  =
\end{equation*}
\begin{equation}\label{S_set_cup_1}
\left({\rm null}\left(\mathbf{H}_{ba}\right)^{\bot}  \cap  {\rm null}\left(\mathbf{H}_{ea}\right)\right)  \cup
\left( {\rm null}\left(\mathbf{H}_{ba}\right)^{\bot} \!\cap\! {\rm null}\left(\mathbf{H}_{ea}\right)^{\bot}\right)  \\
\end{equation}
\begin{equation}\label{S_set_cup_2}
\hspace{-1.5cm} = {\rm null}\left(\mathbf{H}_{ba}\right)^{\bot} \cap \left(  {\rm null}\left(\mathbf{H}_{ea}\right) \cup {\rm null}\left(\mathbf{H}_{ea}\right)^{\bot} \right) \end{equation}
\begin{equation}\label{S_set_cup_3}
\hspace{-6cm} =  {\rm null}\left(\mathbf{H}_{ba}\right)^{\bot}
\end{equation}
\end{subequations}
where (\ref{S_set_cup_2}) and (\ref{S_set_cup_3})  are obtained based on the Distributive Law of sets \cite{Halmos1960book}
and the Rank--Nullity Theorem \cite{Banerjee2014book}, respectively.

From (\ref{Dim_set})--(\ref{S_set_cup}), we obtain
\begin{align} \label{Dim_set_sr}
s + r = {\rm dim}\left(\mathcal{S}_{ba}\right) +  {\rm dim}\left(\mathcal{S}_{be}\right) = {\rm dim}\left( {\rm null}\left(\mathbf{H}_{ba}\right)^{\bot}\right).
\end{align}

Assuming $\mathbf{v}_i \in \mathbb{C}^{N_t \times 1}$ and $\mathbf{u}_j \in \mathbb{C}^{N_r \times 1}$
are the $N_t$ left and $N_r$  right singular vectors of $\mathbf{H}_{ba}$,
respectively, $i = 1,\ldots,N_t$,  $j = 1,\ldots,N_r$, $\mathbf{H}_{ba}$ can be written as
\begin{align} \label{Dim_set_1}
\mathbf{H}_{ba} = \sum\limits_{i = 1}^{{N_1}} {{\lambda _i}{{\bf{u}}_i}{\bf{v}}_i^H}
\end{align}
where ${\lambda _i}$ is the singular value of $\mathbf{H}_{ea}$. For $N_1 < N_t$,
we have
\begin{align} \label{Dim_set_2}
{\rm null}\left(\mathbf{H}_{ba}\right) =  \sum\limits_{i = N_1 + 1}^{{N_t}} {{\omega_i}{{\bf{v}}_i}{\bf{v}}_i^H}
\end{align}
where ${\omega_i}$ denotes an arbitrary non-zero complex value, $i = 1,\ldots,N_t$.
Based on the property of the orthogonal complement of a subspace \cite{Halmos1974book}, we obtain
\begin{subequations}
\begin{equation} \label{Dim_set_3}
\hspace{-2.9cm} \left( {\rm null}\left(\mathbf{H}_{ba}\right)^{\bot}\right)  =  \left(\sum\limits_{i = N_1 + 1}^{{N_t}} {{\omega_i}{{\bf{v}}_i}{\bf{v}}_i^H}\right)^{\bot}
\end{equation}
\begin{equation}
\hspace{-0.2cm} =  {\rm null} \left(\sum\limits_{i = N_1 + 1}^{{N_t}} {{\omega_i}{{\bf{v}}_i}{\bf{v}}_i^H}\right)
\end{equation}
  \begin{equation}
\hspace{-1.8cm} =  \sum\limits_{i = 1}^{{N_1}} {{\omega_i}{{\bf{v}}_i}{\bf{v}}_i^H}.
\end{equation}
\end{subequations}

Therefore, we have
 \begin{align} \label{Dim_set_4}
 {\rm dim}\left( {\rm null}\left(\mathbf{H}_{ba}\right)^{\bot}\right) = N_1.
\end{align}
For $N_1 = N_t$,  ${\rm null}\left(\mathbf{H}_{ba}\right) = \varnothing$, and we obtain
 \begin{align} \label{Dim_set_5}
 {\rm dim}\left( {\rm null}\left(\mathbf{H}_{ba}\right)^{\bot}\right) = N_t.
\end{align}
Combining (\ref{rate}), (\ref{rate_b_GSVD}), (\ref{Dim_set_sr}), (\ref{Dim_set_4}), and (\ref{Dim_set_5}) completes the proof.

\section{Proof of Theorem \ref{PG-GSVD_high}} \label{Proof_PG-GSVD_high}
The key idea of achieving the maximal rate $N_t \log M$ b/s/Hz in the high SNR regime
is to guarantee that all $N_t$ signals can be received by Bob but not by
Eve. To achieve this, $N_s$ signals are combined into a group and transmitted along
the subchannels $\mathbf{R}_r$ in (\ref{eq:sigma_ba_hat}). As a result, we need to analyze
the dimension of $\mathcal{S}_{ba}$.

Based on the Inclusion--Exclusion Principle \cite{Andrews1971book}, we have
\begin{align} \label{Dim_set_2}
{\rm dim}\left(\mathcal{S}_{ba}\right) +  {\rm dim}\left(\mathcal{S}_{n}\right) = {\rm dim} \left(\mathcal{S}_{ba} \cup \mathcal{S}_{n}  \right) -
{\rm dim} \left(\mathcal{S}_{ba} \cap \mathcal{S}_{n}\right).
\end{align}

Following similar steps as in (\ref{S_set_total}) and (\ref{S_set_cup}), we obtain
\begin{subequations}
\begin{equation*}
\hspace{-3cm} \mathcal{S}_{ba} \cap \mathcal{S}_{n}  =
\end{equation*}
\begin{equation} \label{S_set_high}
 \left({\rm null}\left(\mathbf{H}_{ba}\right)^{\bot}  \cap {\rm null}\left(\mathbf{H}_{ea}\right)\right)  \cap
\left( {\rm null}\left(\mathbf{H}_{ba}\right) \cap {\rm null}\left(\mathbf{H}_{ea}\right)\right)
\end{equation}
\begin{equation} \label{eq:Set_2_high}
 = \left(   {\rm null}\left(\mathbf{H}_{ea}\right)  \cap   {\rm null}\left(\mathbf{H}_{ba}\right)^{\bot} \right)  \cap
\left( {\rm null}\left(\mathbf{H}_{ba}\right) \cap {\rm null}\left(\mathbf{H}_{ea}\right)\right)
\end{equation}
\begin{equation} \label{eq:Set_3_high}
\hspace{-0.2cm} =  {\rm null}\left(\mathbf{H}_{ea}\right) \! \cap \! \left(\left( {\rm null}\left(\mathbf{H}_{ba}\right)\right)^{\bot}  \cap {\rm null}\left(\mathbf{H}_{ba}\right)    \right) \! \cap \! {\rm null}\left(\mathbf{H}_{ea}\right)
\end{equation}
\begin{equation} \label{eq:Set_4_high}
\hspace{-7.9cm} = \varnothing
\end{equation}
\end{subequations}
and
\begin{subequations}
\begin{equation*}
\hspace{-3cm} \mathcal{S}_{ba} \cup \mathcal{S}_{n}  =
 \end{equation*}
\begin{equation} \label{S_set_cup_1_high}
\left({\rm null}\left(\mathbf{H}_{ba}\right)^{\bot}   \cap  {\rm null}\left(\mathbf{H}_{ea}\right)\right)  \cup
\left( {\rm null}\left(\mathbf{H}_{ba}\right) \cap {\rm null}\left(\mathbf{H}_{ea}\right) \right)
\end{equation}
\begin{equation} \label{S_set_cup_2_high}
\hspace{-1.5cm}   = {\rm null}\left(\mathbf{H}_{ea}\right) \cap \left(  {\rm null}\left(\mathbf{H}_{ba}\right)^{\bot} \cup {\rm null}\left(\mathbf{H}_{ea}\right) \right)
\end{equation}
\begin{equation}\label{S_set_cup_3_high}
\hspace{-5.8cm} =  {\rm null}\left(\mathbf{H}_{ea}\right).
\end{equation}
\end{subequations}
Since ${\rm rank} \left(\mathbf{H}_{ea}\right) = N_2$,  we have ${\rm dim} \left({\rm null}\left(\mathbf{H}_{ea}\right)\right) = N_t - N_2$.
Then, based on (\ref{Dim_set_2}), (\ref{eq:Set_4_high}), (\ref{S_set_cup_3_high}),  we obtain
\begin{align} \label{Dim_set_3}
r + N_t - k = N_t - N_2.
\end{align}
From (\ref{Dim_set_3}), we know $r = k - N_2$.

When $(k - N_2) N_s \geq N_t$, we design the PG-GSVD precoder in (\ref{eq:precoding_matrix_gsvd}) as follows. We set
\begin{equation}\label{eq:P_high}
\mathbf{P} = \kbordermatrix {~		   &  k -r-s	 & s 		& r 	& N_t - k	\cr
										k - r - s & \mathbf{0} 	& \mathbf{0}  		& \mathbf{0} & \mathbf{0}	\cr
										s                 & \mathbf{0}  	&  \mathbf{0} 	& \mathbf{0} & \mathbf{0}	\cr
										r                 & \mathbf{0}  	& \mathbf{0}      	& {\rm diag}\left(p_1,\ldots p_r\right) 	& \mathbf{0}  \cr
                                       N_t -k                 & \mathbf{0}  	& \mathbf{0}      	& \mathbf{0} 	& \mathbf{0}  \cr  }.
\end{equation}
Also, we select a pairing scheme $\left\{\ell_{1},\ldots,\ell_{N_{t}}\right\}$ in (\ref{eq:P_pair}) satisfying
\begin{equation}\label{eq:pair_high}
\left[\mathbf{P}_s\right]_{ii} = \\
 \left\{ \begin{array}{l}
   0  \quad \quad \quad {\rm if} \  1 \leq i  \leq N_s -1 \\
\frac{p_j}{\omega_{k - r + j}} \ \, {\rm if}  \  i = N_s
 \end{array} \right.
\end{equation}
for $s = 1,\ldots, S$, $i = 1,\ldots, N_s$, and $j = 1,\ldots,r$.

Based on the design in (\ref{eq:P_high}) and (\ref{eq:pair_high}), in the high SNR regime, we have
\begin{align}
 I\left( {{\bf{y}}_{b,s}};\mathbf{x}_{s} \right) & \mathop  \to \limits^{P \to \infty } N_s \log M \label{eq:bob_high}\\
  I\left( {{\bf{y}}_{e,s}};\mathbf{x}_{s} \right)& = 0.  \label{eq:eve_high}
\end{align}

Substituting (\ref{eq:bob_high}) and (\ref{eq:eve_high}) into (\ref{eq:I_pair}) completes the proof.


\end{document}